# Periodically modulated geometric and electronic structure of graphene on Ru(0001)

B. Borca[1], S. Barja[1,2], M. Garnica[1,2], J.J. Hinarejos[1], A.L. Vázquez de Parga[1,2], R. Miranda[1,2], F. Guinea[3]

[1]Dep. Física de la Materia Condensada, Universidad Autónoma de Madrid, Cantoblanco 28049, Madrid, Spain
[2]Instituto Madrileño de Estudios Avanzados en Nanociencia (IMDEA Nanociencia), Cantoblanco 29049, Madrid, Spain
[3]Instituto de Ciencia de Materiales, Consejo Superior de Investigaciones Ciéntificas, Cantoblanco 28049, Madrid, Spain

**Abstract**

We report here on a method to fabricate and characterize highly perfect, *periodically rippled* graphene monolayers and islands, epitaxially grown on single crystal metallic substrates under controlled UHV conditions. The periodicity of the ripples is dictated by the difference in lattice parameters of graphene and substrate, and, thus, it is adjustable. We characterize its perfection at the atomic scale by means of STM and determine its electronic structure in the real space by local tunnelling spectroscopy. There are periodic variations in the geometric and electronic structure of the graphene monolayer. We observe inhomogeneities in the charge distribution, i.e a larger occupied Density Of States at the higher parts of the ripples. Periodically rippled graphene might represent the physical realization of an ordered array of coupled graphene quantum dots. The data show, however, that for rippled graphene on Ru(0001) both the low and the high parts of the ripples are metallic. The fabrication of periodically rippled graphene layers with controllable characteristic length and different bonding interactions with the substrate will allow a systematic experimental test of this fundamental problem.

**1. INTRODUCTION**

The possibility to produce single layers of graphene [1] has opened a fascinating new world of physical phenomena in two dimensions and a new route towards an all-carbon electronics. Graphene has already shown that its charge carriers are massless Dirac fermions [2,3] with large ($1.5 \times 10^4$ cm$^2$/V.s) mobility displaying an anomalous, half-integer integer Quantum Hall Effect [2] even at room temperature [4]. Graphene shows a minimum universal conductance, although with a value approximately $\pi$ times larger than theoretically predicted [2,3]. Systems made up of a few graphene layers can also be grown on a SiC substrate [5]. Recently it has been claimed that free standing isolated graphene layer is intrinsically corrugated [6]. The presence of charge inhomogeneities related to these inherent ripples in graphene sheets has been suggested [7] to explain the larger than predicted value of the universal conductance of graphene, a persistent mystery so far.

Ultra-thin epitaxial films of graphite have been grown, however, on solid surfaces for quite some time [8]. Even the growth of "monolayer-graphite" films onto several substrates by Chemical Vapour Deposition have been reported years ago, but the degree of characterization of the films was hampered by the experimental limitations existing at that time [9]. In the last two years several groups have taken advantage of local probe techniques to study the geometric and electronic structure of graphene epitaxially grown on different metallic substrates such as Ir(111) [10-12], Pt(111) [13,14], Ni(111) [15], Ru(0001) [16-18]. Graphene grown on metallic substrates could be important as a practical source of large graphene samples after dissolving the metallic substrate, but it is also important from a fundamental point of view. Depending on the metallic substrate, the bonding interaction with the graphene monolayer, and, accordingly, the modification, of its electronic structure is different [19]. For systems such as Ir(111), the graphene electronic structure is believed to be only weakly modified by the metallic substrate [12], while for others such as Ru(0001) it is strongly modulated by the interaction with the metal [16]. This interaction can be modified by the intercalation of another metal between the graphene layer and the metallic substrate [20,21]

## 2. EXPERIMENTAL RESULTS

The samples were prepared in an Ultra High Vacuum (UHV) chamber with a base pressure in the range of $10^{-11}$ Torr. The chamber is equipped with a variable temperature scanning tunnelling microscope (STM), a Low Energy Electron Diffraction (LEED) optics that allow us to do Auger spectroscopy, and facilities for ion sputtering , evaporation and gas exposure. The polycrystalline tips were cleaned *in situ* by $Ar^+$ sputtering at 2.5kV and annealing by electron bombardment to prepare a tip clean and blunt for the spectroscopic measurements [22]

The substrate is a single crystal of Ru exposing the (0001) surface, which was cleaned in UHV by ion sputtering and annealing to 1400 K [23]. Atomically resolved STM images of the clean surface show the triangular lattice of Ru atoms separated 0.27 nm, while STS only reveals the $p_z$-like surface state of Ru(0001) crossing the Fermi energy [23]. The graphene samples are grown ) under UHV conditions either by controlled segregation of C from the bulk of the substrate, by thermal decomposition at 1000 K of ethylene molecules pre-adsorbed at 300 K on the sample surface or by exposing the Ru(0001) surface, held at 1000 K, to a ethylene partial pressure of $5 \times 10^{-7}$ Torr for several minutes. Depending on the amount of ethylene, a continuous graphene film or nanometer-sized islands can be produced on the surface [16]. The epitaxial graphene monolayer covers uniformly the Ru substrate over lateral distances larger than several microns, faithfully reproducing the atomic steps, screw dislocations and other structural defects of the Ru substrate (see Figs. 1a). Low Energy Electron Microscopy (LEEM) images have revealed that the graphene layer extends for more than 100 microns [17].

The graphene monolayer is easily recognized by the presence of a triangular array of bumps with an average separation of around 3nm (2.93±0.08 nm) [16]. The presence of these bumps can be understood in a simple approach as originating from a Moiré pattern due to the difference between the lattice parameter of the carbon overlayer and the Ruthenium surface. If an unperturbed graphene layer is placed on top of a Ru(0001) surface, the lattice of graphene will be incommensurate with the underlying lattice of Ru. The lattice parameter relation implies that 11 carbon honeycombs (2.707 nm) will adjust almost exactly to 10 Ru-Ru interatomic distances (2.706 nm), with a negligible compressive strain of 0.05%. It should be noticed that a (12x12) superstructure, i.e. 12 C-C (2.953 nm) vs. 11 Ru-Ru ones (2.976 nm) will be only under 0.78% tensile strain.  Early work by Grant and Haas [24] described a (9x9) superstructure due to C segregation from the bulk of Ru, while Goodman et al [25] described a (11x11) superstructure after decomposing methane and heating to 1300K. Marchini et al. [26] proposed, based on STM and LEED measurements, a (12x12) superstructure that gives periodicity between ripples of 3nm. In a recent work, Martoccia et al [27], based on a fit to Surface X-ray Diffraction data, propose a larger (25×25) supercell that takes into account the small bucklings (0.02 nm) and lateral relaxations (0.01nm) induced *in* the Ru(0001) surface by the graphene overlayer.

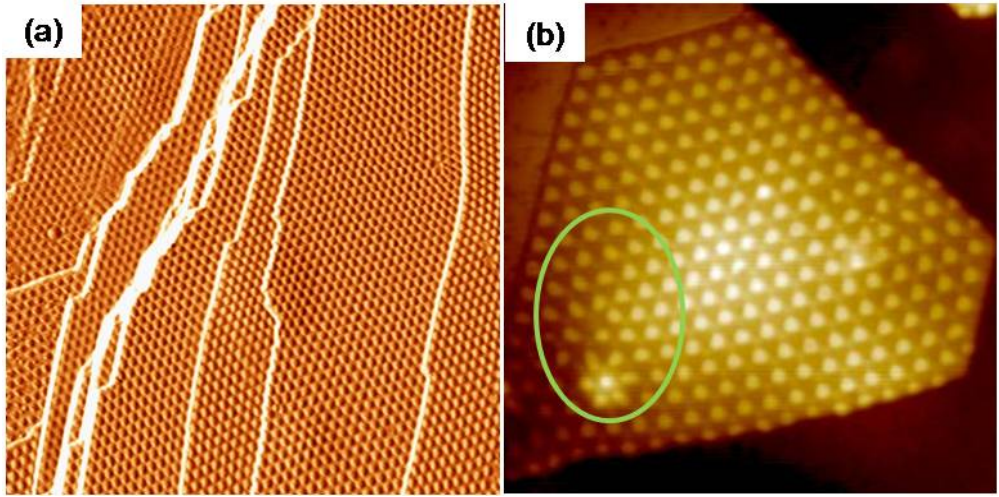

**Figure 1:** *(a) 150 nm × 150 nm STM image of a graphene monolayer covering completely the Ru(0001) surface. (b) 80 nm × 80 nm STM image of the Ru(0001) surface partially covered with monolayer-high graphene islands. The green circles mark areas of the islands where the presence of defects modifies the periodicity of the Moiré pattern. The image in panel a have been differentiated along the X-direction [37].*

This large surface unit cell makes very difficult a detailed calculation of the surface electronic structure. So far two different groups have reported Density Functional Theory (DFT) calculations on this system. In both cases the Ruthenium substrate is simulated with only three atomic layers. Wang et al. [28] did a calculation for a 3nm periodic structure. They impose a unit cell with 12 carbon atoms every 11 Ruthenium ones and they found a corrugated graphene layer with a minimum distance with the Ru substrate of 0.22 nm and a substantial geometrical corrugation of 0.16 nm. They also founded carbon atoms with different bonding interactions with the substrate. The carbon atoms located at the higher areas of the superstructure are on three-fold hollow sites and have almost no interaction with the Ru atoms. On the contrary, some of the carbon atoms closer to the Ru substrate occupy top positions and present a strong chemical bonding with the substrate. Jiang et al. [29] tried two different graphene superstructures with 3.0 and 2.7 nm periodicities. For the one with a periodicity of 3 nm, they found an adhesion energy of 21 meV/C atom, in agreement with Wang et al [28], as well as a 3% stretching of the C-C bonds for the atoms located at the lower region. The compressive strain produced in the rest of the graphene layer produces the periodic bucking [29]. As mentioned before, though, the lowest strain between the carbon and Ruthenium lattice corresponds to the (11x11) superstructure with a periodicity of 2.7 nm. In spite of that reduced strain energy, the increased C-Ru bonding in the superstructure with a periodicity of 3nm results in a cohesion energy 10 meV/C atom larger. According to this calculation, the balance of strain and bonding results in the (12x12) being energetically more stable [29].

Upper panel in Figure 2 shows a high resolution topographic STM image of the epitaxial graphene layer, revealing its atomic structure superimposed on the modulated superstructure of the Moiré pattern. Unlike STM images of graphite [30], which show normally only one of the two C atoms in the surface unit cell, the honeycomb structure of graphene is clearly resolved in the upper part of the ripples with its 0.14 nm C-C distance. The lower panel in Figure 2 shows two line profiles measured on the atomically resolved images indicating that the apparent atomic corrugation of the carbon atoms do not depend on the position with respect to the ripples.

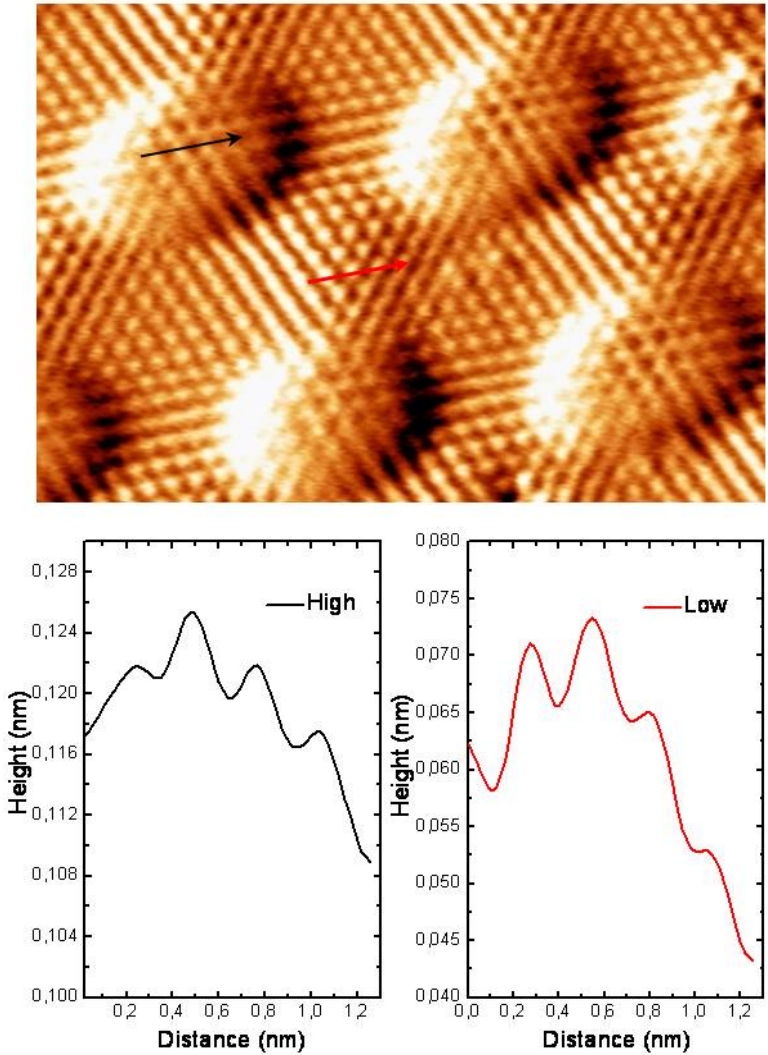

**Figure 2:** *(a) Atomically resolved STM image showing one unit cell of the Moiré pattern. The image has been differentiated along the X-direction The lower panels show two line profiles measured on the upper and lower parts of the Moiré pattern. The atomic corrugation in both profiles is of the order of 5 pm, i.e. it does not depend on which area of the Moiré pattern is measured. The image was measured with a sample bias voltage of -5mV and a tunnelling current of 3nA.*

The presence of steps or other defects on the Ru(0001) surface reduces slightly the periodicity of the graphene ripples in the Moiré pattern. One example of this is shown on Figure 1(b) where a defect can be seen on the graphene islands (green circle). The presence of the defect modifies up to 15% the distance between the ripples around it. This result is in agreement with the calculations by Jiang et al. [29], where two periodicities were found to co-exist, with the larger (3.0 nm) being more stable than the other one (2.7 nm). The presence of a defect modifies the delicate energy balance between bonding with the substrate and strain due to the different lattice parameter.

We use the islands to measure the distance between the graphene layer and the Ru surface. Figure 3(a) shows the STM image of one of these islands and Figure 3(b) two profiles measured at +0.8V and -0.8V on the island edge. The apparent vertical corrugation of the rippled graphene monolayer, as seen with STM, changes with the tunnelling voltage. Figure 3(c) shows the evolution of the apparent height of the upper (red dots) and lower (black dots) areas of the

Moiré. While the upper area reduces its apparent height when the images are measured at positive bias voltage the apparent height of the lower areas remains almost constant around 0.145nmm as demonstrated in the profile shown in Figure 3(b). This is much smaller than the interlayer distances in graphite (0.335 nm), indicating strong C-Ru bonding at the lower parts of the superstructure, again in line with DFT calculations [28,29] predicting a C-Ru distance of 0.22 nm. It is also remarkable that the periodically rippled coincidence lattice goes right to the steps of the islands. In fact in some cases (e.g. see the lower edge of the island) the island step cuts through the (12x12) arrangement.

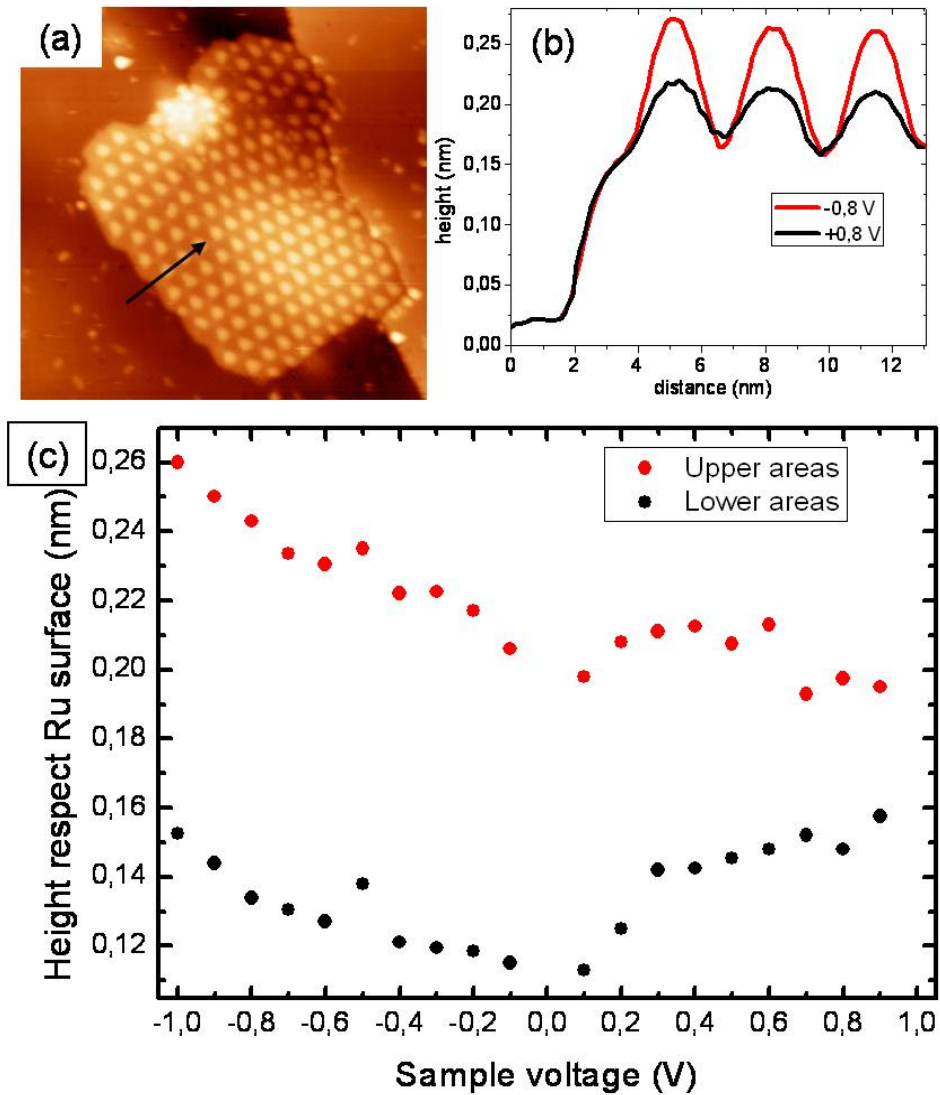

**Figure 3:** *(a) STM topography of a graphene island attached to an atomic step of the Ru(0001) surface. (b) Line profiles measured at positive (black) and negative (red) sample bias voltage. (c) Evolution of the apparent height of the upper (red dots) and lower (black dots) areas of the Moiré pattern with the sample bias voltage.*

Figure 4(a) shows a compilation of the apparent corrugation of the graphene Moiré pattern measured with different tips on different samples covered with a continuous graphene layer. As can be seen the corrugation values go from 0.1 nm (at a sample voltage of –1 V) down to 0.02 nm (at +2 V) in agreement with the data measured on the samples covered with graphene islands. If these apparent corrugation values are compared with the values given by the DFT calculations [28, 29] it is clear that the electronic effects are extremely important in this system. Furthermore, if the tunnelling sample voltage is increased above +2.8V, there is an inversion of the contrast as seen in Figure 4(b). The defect and the line profile are marked as a guide for the eye. This means that the electronic effect is able to compensate the surface corrugation that si calculated to be of the order of 0.16 nm [28, 29].

The bonding with the substrate occurs through the hybridization of the C π states with the Ru d states as shown by the DFT calculations. Photoelectron spectroscopy shows clearly the presence of two types of carbon atoms [19]. Angular Resolved Photo Electron Spectroscopy (ARPES) data show that the graphene bands in graphene/Ru(0001) are similar to the ones of graphite, but rigidly shifted down in energy [31]. The bottom of the π band at the centre of the Brillouin Zone is shifted down by 1.8 eV with respect to graphite [31]. The small energy shift of the C1s core level with respect to graphite indicates that the charge transfer from the substrate is small, but not negligible, i.e. the graphene layer is doped with electrons from the substrate.

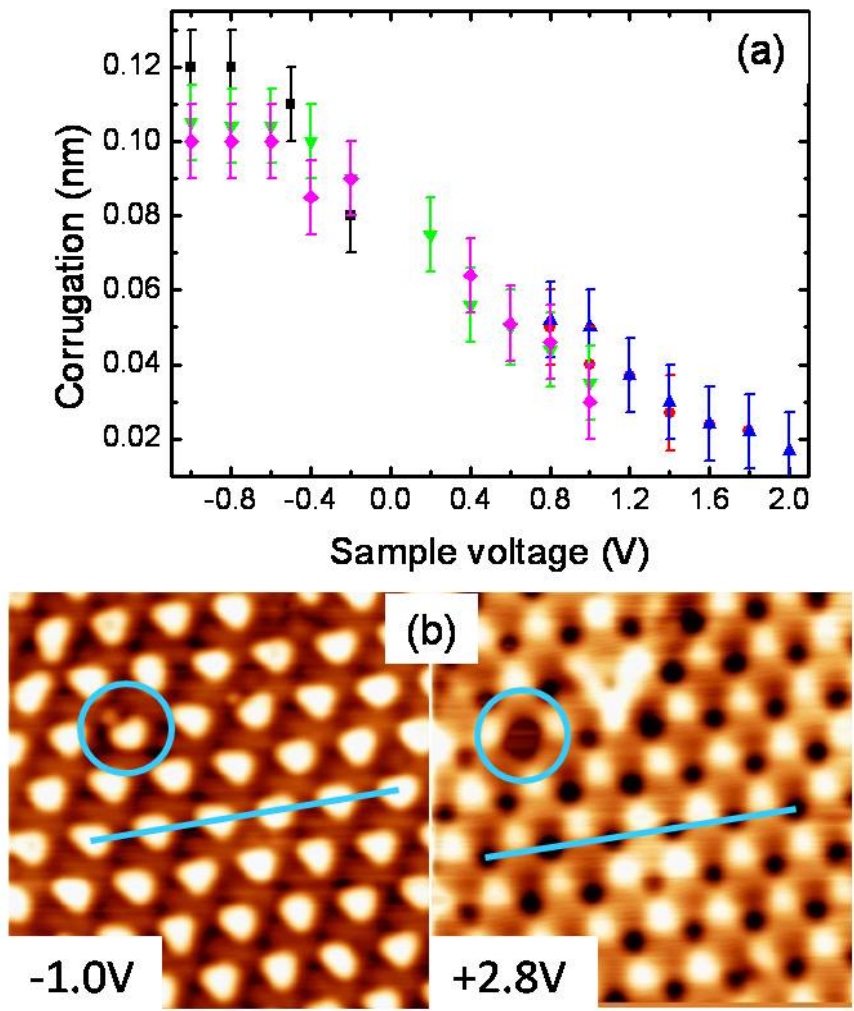

**Figure 4:** *(a) Evolution of the apparent corrugation of the Moiré pattern with the sample bias voltage. The measurements have been performed with several tips and samples. (b) "Topographic" images measured simultaneously at negative (left) and positive (right) sample bias voltage. The blue circle marks a defect present in the Moiré that it is used to compare the two images. For sample bias voltage close to +3 V the apparent corrugation of the Moiré pattern is inverted. The blue line is a guide for the eye.*

Figure 5(a) shows (dI/dV vs V) tunnelling spectra, which are roughly proportional to the Local Density of States (LDOS), recorded at 300 K on top of the "upper" and "lower" regions of the corrugated graphene layer. The experimental tunnelling spectra recorded at different spatial positions are obviously different: the occupied LDOS is systematically larger in the "high" areas of the rippled layer, while the empty LDOS is larger in the "low" parts. The differences are robust enough to survive at 300 K. The periodic charge inhomogeneities in the graphene layer can be visualized *directly* in the real space by imaging the spatial distribution of dI/dV close to the Fermi energy [16]. This modulation in the electronic structure of the graphene layer is stable even for very small graphene nanoislands. In Figure 5(b) we show the edge of a graphene layer with the spatially resolved tunnelling spectra measured simultaneously for the occupied (lower panel) and unoccupied (upper panel) LDOS. As can be seen in the figure the occupied LDOS is larger at the "upper" areas of the superlattice and smaller in the lower areas

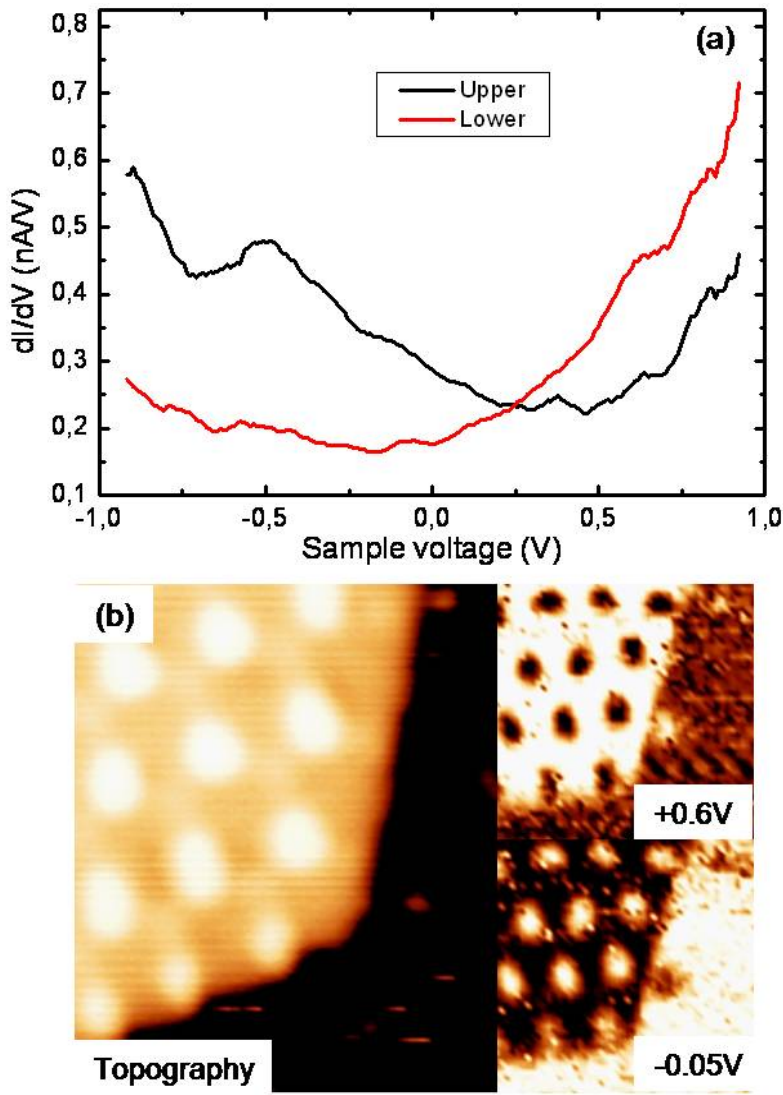

**Figure 5:** *(a) Tunnelling spectra of extended graphene/Ru(0001) recorded at 300 K on the upper (black) and lower (red) areas of the graphene Moiré pattern. (b) Spatial map of the tunnelling conductance (dI/dV) on the edge of a graphene island. The dI/dV map measured at -50meV shows the presence of electron pockets intact even at the edge of the island. The dI/dV map measured at positive sample bias voltage shows that the unoccupied LDOS is higher at the lower areas of the Moiré superstructure.*

## 3. THEORETICAL MODEL

In order to understand the physics behind these measurements we use a very simple model that contains the main features of the system. The electronic structure has been simulated by assuming that the rippled graphene layer is not too distorted. The model calculations have been performed for an isolated graphene layer in which the effect of the substrate has been considered to result in: i) a shift of the Dirac point by -0.3 eV due to doping; ii) the introduction of a finite lifetime caused by hybrization of the π orbitals with the band of the substrate; and iii) a (12x12) periodic potential that changes between -3V and 3V/2, where V=-0.3eV to account for the periodic structural ripples. The π band in graphene has a total width of W~6t, where t ~3 eV is the hopping between π orbitals at nearest neighbour C atoms [16].

We do not include modulations in the hoppings, or topological lattice defects, which can lead to an effective gauge potential [32], as the main effect of the superstructure is the transfer of spectral weight between the valence and conduction bands.

The shift of the electronic levels is given by:

$$V(x,y) = V_0 \left[ \cos\left(\frac{\pi x}{5\sqrt{3}a} + \frac{\pi y}{15a}\right) + \cos\left(\frac{2\pi y}{15a}\right) + \cos\left(-\frac{\pi x}{5\sqrt{3}a} + \frac{\pi y}{15a}\right) \right]$$

Where a=1.4 Angstroms is the distance between carbon atoms. The resulting potential V(x,y), for $V_0$=0.3eV is shown in Figure 6.

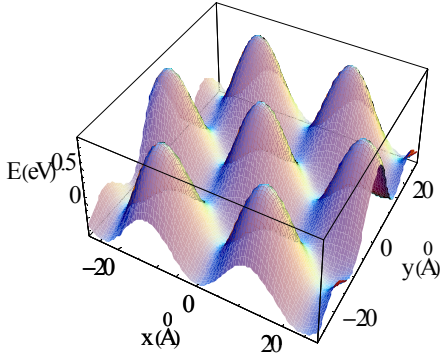

**Figure 6:** *Three dimensional plot of the potential induced by the graphene superstructure.*

In order to calculate the DOS, we use a (30x30) unit cell, and sum over 6 special points in the irreducible sector of the Brillouin Zone [32]. Hence, the total number of states included in the calculation is 10800. This is sufficient to resolve local changes in the electron density in the order $10^{-2} – 10^{-3}$. The accuracy of the calculation can be appreciated from Figure 7, where the calculated total DOS of unperturbed graphene is shown.

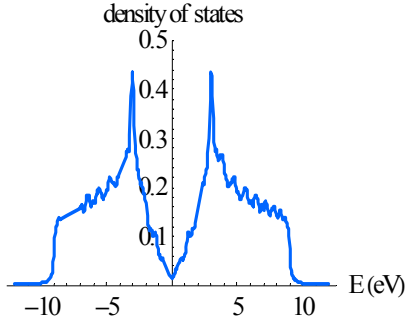

**Figure 7:** *Calculated DOS of unperturbed graphene. The levels have a finite broadening of 0.1eV. The resonances above the van Hove singularities are due to finite size effects. The nature of the subbands can be inferred by comparing calculations obtained with different broadenings.*

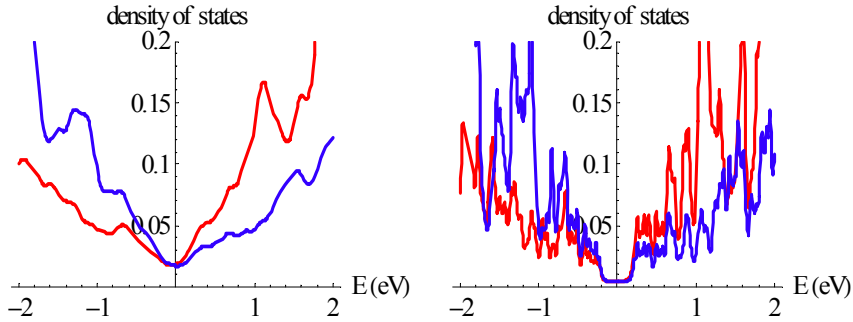

**Figure 8:** *DOS near the Dirac energy calculated with different broadenings. Left: 0.1eV. Right: 0.02eV. Red) and blue plots give the density of states at different positions in the unit cell of the superlattice.*

The superstructure considered here does not superimpose the two inequivalent corners of the Brillouin Zone, and it does not open a gap at the Dirac energy [33]. The spectrum is splitted in subbands separated by gaps, away from the K and K' points. That is shown in Figure 8. This comparison has been done using a (20x20) supercell. The plots with the smallest broadening, 0.02eV, allow us to resolve the individual levels included in the calculation. Their clustering at certain energies is due to the formation of subbands. The calculation with a broadening more consistent with the experimental resolution show peaks with correspond with the position of the subbands, and are independent of finite size effects.

We finally show in Figure 9 the dependence of the density of states near the Dirac energy on the strength of the potential induced by the superlattice. In all cases, there is a transfer of DOS from the conduction to the valence band in some atoms, and the opposite in the others.

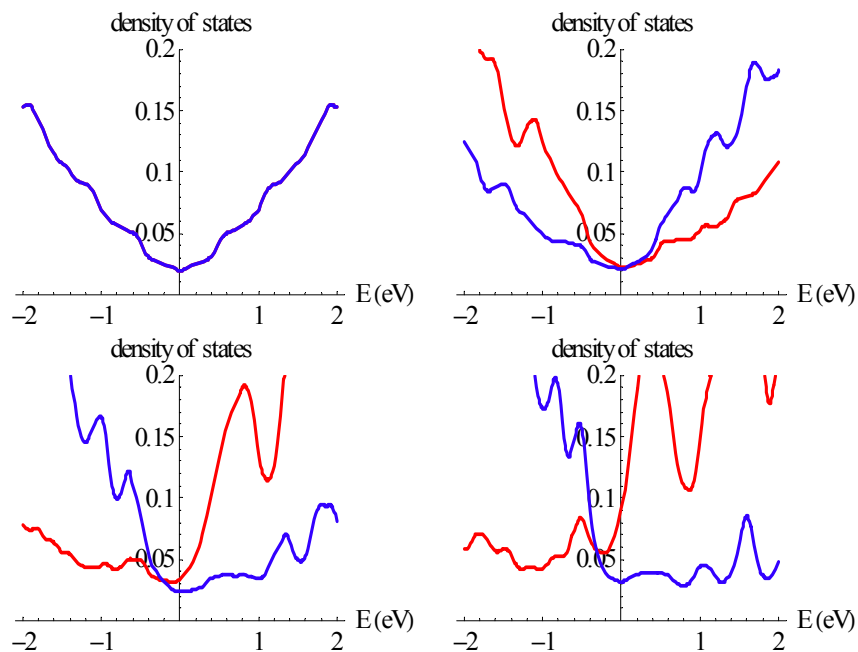

**Figure 9:** *Density of states for different values of the superlattice potential. Top left: $V_0=0$. Top right: $V_0=0.3eV$. Bottom left: $V_0=0.6eV$. Bottom right: $V_0=0.9eV$.*

## 4. CONCLUSIONS

In agreement with the experiments, the calculations show that the occupied LDOS is larger on the "high" regions of the superlattice, where the potential is at a minimum, while the empty LDOS is larger at the "low" regions of the graphene layer. This effect is extremely robust and indicates that the valence band is depleted in the low portions of the ripples, while the conduction band is depleted at the high parts of the corrugated graphene layer. The potential $V_0=0.3eV$ gives results in better agreement with the experiments [16]. The calculations also reproduce the weaker features at both sides of the Fermi energy, and reveal that they reflect the existence of the periodic superlattice, which induces a folding of the graphene bands in the new (12x12) Brillouin Zone.

On a poor metallic (or semimetallic), 2D graphene monolayer, the lattice misfit and its periodic modulation of the bonding with the metallic substrate originates both a geometric and an electronic modulation. The electronic inhomogeneities are reflected in a larger occupied DOS close to the Fermi energy at the upper part of the structural ripples. Both the upper and the lower parts of the structural ripples are, however, weakly metallic or semimetallic. Thus, the inhomogeneities do not translate into the formation of isolated Quantum Well, but they can originate a periodic array of preferential adsorption sites for donor/acceptor molecules, a topic of the utmost importance for nanostructuring functional molecular layers [35].

The methods described here can be implemented on many other single crystal substrates, giving rise to a series of graphene monolayers with different, substrate-dependent, periodic corrugations and, thus, opening the possibility to systematically test the electronic properties of controlled, charge inhomogeneous graphene layers. The opening of a series of minigaps [36] in the graphene bands by the additional periodic potential of the superlattice is also expected to give rise to new phenomena at low temperatures in the presence of high magnetic fields.

**ACKNOWLEDGEMENTS**

Partial financial support by the Ministerio de Ciencia y Tecnología through projects FIS2005-05478-C02-01 and CONSOLIDER "Nanociencia molecular" CSD2007-00010, the Comunidad de Madrid, through the programs CITECNOMIK CM2006-S0505/ESP/0337 and NANOMAGNET S0505/MAT/0194.

**REFERENCES**

[1] K.S. Novoselov, A.K. Geim, S.V. Mozorov, D. Jiang, Y. Zhang, S.V. Dubonos, I.V. Grigorieva and A.A. Firsov, *Science* **306,** 666 (2004); K.S. Novoselov, D. Jiang, F. Schedin, T.J. Booth, V.V. Khotkevich, S.V. Morozov and A.K. Geim, Proc. Nat. Acad. Sci. **102**, 10451 (2005); A. K. Geim and K. S. Novoselov, *Nature Materials* **6**, 183 (2007).
[2] K.S. Novoselov, A.K. Geim, S.V. Mozorov, D. Jiang, M.I. Katsnelson, I.V. Grigorieva, S.V. Dubonos and A.A. Firsov, *Nature* **438**, 197 (2005); Y. Zhang, Y.-W. Tan, H.L. Stormer am P. Kim, *Nature*, **438**, 201-204 (2005); K.S. Novolesov, E. McCann, S.V. Morozov, V.I. Falko, M.I. Katsnelson, U. Zeitler, D. Jiang, F. Schedin and A.K. Geim, *Nature Physics* **2**, 177-180 (2006).
[3] V. P. Gusynin and S. G. Sharapov, *Phys. Rev. Lett*. **95**, 146801 (2005). N. M. R. Peres, F. Guinea and A. H. Castro Neto, *Phys. Rev. B* **73**, 125411 (2006).
[4] K.S. Novoselov, Z. Jiang, Y.Zhang, S.V. Morozov, H.L. Stormer, U. Zeitler, J.C. Maan, G.S. Boebinger, P. Kim and A.K. Geim, *Science*, **315**, 1379 (2007).
[5] C. Berger, Z. Song, T. Li, X. Li, A.Y. Ogbazghi, R. Feng, Z. Dai, A.N. Marchenkov, E.H. Conrad, P.N. First and W.A. de Heer, *J. Phys. Chem. B* **108**, 19912 (2004).
[6] J.C. Meyer, A.K. Geim, M.I. Katsnelson, K.S. Novoselov, T.J. Booth and S. Roth, *Nature*, **446**, 60 (2007).
[7] J. Martin, N. Kerman. G. Ulbricht, T. lohmann, J.H. Smet, K. von Klitzing and A. Yacobi, Nature Phys. **4**, 144 (2008); S. Cho and M. S. Fuhrer, "Charge Transport and Inhomogeneity near the Charge Neutrality Point in Graphene", arXiv:0705.3239 (2007); E. H. Hwang, S. Adam and S. Das Sarma, *Phys. Rev. Lett.*, **98**, 186806 (2007).
[8] For a review see Ch. Oshima and A. Nagashima, *J. Phys.: Condens. Matter* **9**, 1 (1997).
[9] Z-P Hu, D.F. Ogletree, M.A. van Hove and G.A. Somorjai, *Surface Sci*. **180**, 433 (1987).
[10] A.T. N'Diaye, J. Coraux, T.N. Plasa, C. Busse and T. Michely, *New J. Phys*. **10**, 043033 (2008).
[11] J. Coraux, A.T. N'Diaye, M. Engler, C. Busse, D. Wall, N. Buckanie, F.-J. Meyer zu Heringdorf, R. van Gastel, B. Poelsema and T. Michely, *New J. Phys*. **11** 023006 (2009)
[12] I. Pletikosic, M. Kralj, P. Pervan, R. Brako, J. Coraux, A.T. N'Diaye, C. Busse, T. Michely, *Phys. Rev. Lett.* **102**, 056808 (2009)
[13] T.A. Land, T. Michely, R.J. Behm, J.C. Hemminger and G. Comsa, *Surf. Sci.***264**, 261 (1992)
[14] H. Ueta, M. Saida, C. Nakai, Y. Yamada, M. Sasaki, and S. Yamamoto, *Surf. Sci.* **560**, 183 (2004)
[15] Y.S. Dedkov, M. Fonin, U. Rüdinger and C. Laubschat, *Phys. Rev. Lett.* **100**, 107602 (2008)
[16] A.L. Vázquez de Parga, F. Calleja, B. Borca, M.C.G. Passeggi, J.J. Hinarejos, F. Guinea and R. Miranda, *Phys. Rev. Lett.* **100**, 106802 (2008).
[17] P.W. Sutter, J.-I. Flege and E.A. Sutter, *Nat. Mater.* **7**, 406 (2008)
[18] T. Brugger, S. Günther, B. Wang, H. Dil, M.-L. Bocquet, J. Osterwalder, J. Wintterlin and T. Greber, *Phys. Rev. B* **79**, 045407 (2009)
[19] A.B. Preobrajenski, M.L. Ng, A.S. Vinogradov and N. Måtersson, *Phys. Rev. B* **78**, 073401 (2008)
[20] A.M. Shikin, D. Farías, V.K. Adamchuk and K.-H. Rieder, *Surf. Sci.* **424**, 155 (1999)
[21] A. Varykhalov, J. Sánchez-Barriga, A.M. Shikin, C. Biswas, E. Vescovo, A. Rybkin, D. Marchenko and O. Rader, *Phys. Rev. Lett.* **101**, 157601 (2008)
[22] A.L. Vázquez de Parga, O.S. Hernán, R. Miranda, A. Levy Yeyati, N. Mingo, A. Martín-Rodero and F. Flores, *Phys. Rev. Lett.* **80**, 357 (1998)

[23] F. Calleja, A. Arnau, J.J. Hinarejos, A.L. Vázquez de Parga, W. Hofer, P.M. Echenique and R. Miranda, *Phys. Rev. Lett.* **92**, 206101 (2004).
[24] J.T. Grant and T.W. Haas, Surf. Sci. **21**, 76 (1970).
[25] M.C. Wu, Q. Xu and D. W. Goodman, J. Phys. Chem. **98**, 5104 (1994).
[26] S. Marchini, S. Günther and J. Wintterlin, *Phys. Rev. B* **76**, 075429 (2007)
[27] D. Martoccia, P.R. Willmott, T. Brugger, M. Björck, S. Günther, C.M. Schlepütz, A. Cervellino, S.A. Pauli, B.D. Patterson, S. Marchini, J. Wintterlin, W. Moritz and T. Greber, *Phys. Rev. Lett.* **101**, 126102 (2008)
[28] B. Wang, M.-L. Bocquet, S. Marchini, S. Günther and J. Wintterlin, *Phys. Chem.Chem. Phys.* **10**, 3530 (2008)
[29] D. Jiang, M.-H. Du and S. Dai, *J. Chem. Phys.* **130** 074705 (2009)
[30] O.V. Sinitsyna and I.V. Yaminski, *Russian Chem. Review* **75**, 23 (2006)
[31] F.J. Himpsel, K. Christmann, P. Heimann and D.E. Eastman, *Surf. Sci.* **115**, L159 (1982)
[32] J. Gónzalez, F. Guinea and M.A.H. Vozmediano, *Phys. Rev. Lett.* **69**. 172 (1992); S.V. Mozoroz, K.S. Novoselov, M.I. Katsnelson, F. Schedin, L.A. Ponomarenko, D. Jiang and A.K. Geim, *Phys. Rev. Lett.* **97** 016801 (2006); A. Morpugo and F. Guinea *Phys. Rev. Lett.* **97**, 196804 (2006)
[33] D.J. Chadi and M.L. Cohen, *Phys. Rev. B* **8**, 5747 (1973); S.L. Cunningham, *Phys. Rev. B* **10**, 4988 (1974).
[34] J.L. Mañes, F. Guinea and M.A.H. Vozmediano, *Phys. Rev. B* **75**, 155424 (2007)
[35] R. Otero, D. Ecija, G. Fernandez, J.M. Gallego, L. Sánchez, N. Martín and R. Miranda, Nanolett. **7**, 2602 (2007)
[36] Gaps like related to the SiC substrate have been observed in graphene samples prepared by vaccum evaporation of SiC, see V.W. Brar, Y. Zhang, Y. Yayon, T. Ohta, J.L. MacChesney, A. Bostwick, E. Rotengber, K. Horn and M.F. Crommie, Appl. Phys. Lett. **91**, 122102 (2007)
[37] I. Horcas et al., Rev. Sci. Instrum. **78**, 013705 (2007)